\documentclass[twocolumn,aps,prd,preprintnumbers,showpacs,superscriptaddress,nofootinbib,amsmath,amssymb,floats,floatfix,showkeys,notitlepage,longbibliography]{revtex4-1}
\usepackage{comment}
\usepackage{graphicx}
\usepackage{subfigure}
\usepackage{palatino}
\usepackage[commandnameprefix=always]{changes}
\usepackage{hyperref}
\hypersetup{colorlinks=true,linkcolor=blue,urlcolor=blue,citecolor=blue}
\usepackage[toc,page]{appendix}
\usepackage[normalem]{ulem}

\usepackage{lipsum}
\usepackage{graphicx}
\usepackage{subfigure}
\usepackage{palatino}
\usepackage{sans}
\usepackage{adjustbox}
\usepackage{latexsym}
\usepackage{amsmath}
\usepackage{amssymb}
\usepackage{amsfonts}
\usepackage{dcolumn}
\usepackage{bm}
\usepackage{tikz}
\usepackage{bigints}
\usepackage{array,tabularx,multirow,booktabs}
\usepackage[tracking=true]{microtype}
\SetTracking{}{500}
\SetTracking{encoding={*}, shape=sc}{40}
\allowdisplaybreaks
\usepackage{adjustbox}
\usepackage{latexsym}
\usepackage{amsmath}
\usepackage{amssymb}
\usepackage{amsfonts}
\usepackage{dcolumn}
\usepackage{bm}
\usepackage{tikz}
\usepackage{bigints}
\usepackage{array,tabularx,multirow,booktabs}
\usepackage[tracking=true]{microtype}
\usepackage{color}
\allowdisplaybreaks

\begin{document} \sloppy

\title{Nonlinear Einstein-Power-Yang-Mills AdS Black Holes: From Quantum Tunneling to Aschenbach Effect}

\author{Erdem Sucu
}
\email{erdemsc07@gmail.com}
\affiliation{Physics Department, Eastern Mediterranean
University, Famagusta, 99628 North Cyprus, via Mersin 10, Turkiye}

\author{\.{I}zzet Sakall{\i}} 
\email{izzet.sakalli@emu.edu.tr}
\affiliation{Physics Department, Eastern Mediterranean
University, Famagusta, 99628 North Cyprus, via Mersin 10, Turkiye}

\begin{abstract}
This study investigates the thermodynamic and quantum properties of Einstein-Power-Yang-Mills (EPYM) black holes in an Anti-de Sitter background, focusing on the effects of the nonlinear Yang-Mills charge parameter $\gamma$. We derive the metric function, analyze Hawking radiation through boson tunneling, and calculate thermodynamic properties including temperature and phase transitions. The quantum tunneling of $W^+$ bosons is examined using the WKB approximation and Hamilton-Jacobi formalism, revealing how nonlinearity modifies the radiation spectrum. We compute the effective potential governing photon orbits and null geodesics, demonstrating significant alterations in light behavior in strong gravitational fields. Additionally, we explore the Aschenbach effect, showing that this phenomenon, which is typically associated with rotating black holes, can emerge in spherically symmetric EPYM spacetimes because of non-linear field interactions. Our results may yield observational markers that can be identified with instruments such as the Event Horizon Telescope and upcoming gravitational wave detectors.
\end{abstract}

\date{\today}


\maketitle
\section{Introduction}

Black holes represent some of the most fascinating objects in theoretical physics, serving as crucial testing grounds for our understanding of gravity in extreme conditions. These enigmatic entities, predicted by Einstein's general relativity, continue to challenge our comprehension of fundamental physics by bringing together concepts from quantum mechanics, thermodynamics, and high-energy physics. Among the various theoretical frameworks developed to study black holes, the EPYM model has emerged as particularly significant, offering important insights into how nonlinear gauge fields affect spacetime geometry and related physical phenomena.

The EPYM theory extends the standard Einstein-Maxwell framework by incorporating nonlinear Yang-Mills (YM) fields characterized by a power parameter $\gamma$, which controls the degree of nonlinearity in electromagnetic interactions. This nonlinearity substantially modifies the black hole solution, introducing distinctive features not present in linear theories \cite{mazharimousavi2009lovelock,stetsko2021static,mazharimousavi2011black}. The action for the EPYM model typically takes the form
\begin{equation} \label{izaction}
 \mathcal{I} = \frac{1}{2}\int d^4x \sqrt{-g}(R - 2\Lambda - [Tr(F^{(a)}_{\mu\nu}F^{(a)\mu\nu})]^{\gamma}),   
\end{equation}
where $R$ is the Ricci scalar, $\Lambda$ is the cosmological constant, and the parameter $\gamma$ determines the strength of the nonlinearity in the YM field. In Anti-de Sitter (AdS) backgrounds, these nonlinear effects become particularly pronounced and lead to rich thermodynamic behaviors including multiple phase transitions and critical phenomena \cite{hawking1983thermodynamics,ghosh2019contact}.

The thermodynamic properties of EPYM AdS black holes have attracted considerable attention in recent years, particularly within the context of gauge-gravity duality and the AdS/CFT correspondence \cite{mazharimousavi2009lovelock,mazharimousavi2011black}. Research has shown that these black holes exhibit complex phase structures reminiscent of van der Waals fluids, with the nonlinearity parameter $\gamma$ playing a crucial role in determining the critical behavior. Such studies have revealed that EPYM black holes can undergo phase transitions between small and large black hole states, with potential implications for understanding strongly coupled quantum field theories through holographic principles \cite{wei2013critical,kubizvnak2012p}.

Beyond their thermodynamic characteristics, EPYM black holes also exhibit intriguing quantum features, particularly with regard to Hawking radiation and information paradox considerations. The quantum tunneling of particles from the black hole horizon represents a semiclassical approach to understanding Hawking radiation \cite{nozari2008hawking,erbin2018universality,jiang2006hawking,chen2008hawking,jusufi2018quantum,sakalli2015tunnelling,sucu2023gup}. For EPYM black holes, this process becomes especially complex as a result of the influence of the non-linear YM field on the spacetime geometry. By applying the WKB approximation and the Hamilton-Jacobi formalism, the tunneling probability can be derived for various particles, including scalar particles, fermions, and vector bosons such as $W^+$ \cite{li2016massive,li2015massive,gecim2018gup,javed2019tunneling,gecim2017massive,jusufi2016tunneling,gecim2018quantum}. The tunneling rate, related to the imaginary part of the action across the horizon, yields a temperature consistent with the standard thermodynamic definition but modified by nonlinear effects \cite{akhmedov2006hawking,pilling2008black}.

The effective potential governing particle motion around EPYM black holes provides another window into their unique physical properties. This potential, derived from the geodesic equations or the Klein-Gordon equation for test fields, characterizes how particles and fields propagate in the black hole spacetime \cite{sadeghian2021separability,frolov2007separability}. For EPYM black holes, the effective potential exhibits distinctive features related to the nonlinearity parameter $\gamma$, affecting stability regions, particle confinement, and radiation emission profiles. Understanding an effective potential is crucial for predicting observational signatures such as gravitational wave emission, accretion disk properties, and quasinormal modes \cite{ferrari2008quasi,nagar2007accretion}.

Photon orbits and null geodesics around black holes have become increasingly important with recent advancements in observational astrophysics, particularly with the Event Horizon Telescope's imaging of supermassive black hole shadows \cite{sucu2025dynamics}. In EPYM black holes, these null geodesics exhibit unique characteristics influenced by the nonlinear YM field. The photon sphere, where light can orbit the black hole in circular paths, plays a critical role in determining the black hole's shadow and lensing properties \cite{sucu2024effect}. For EPYM black holes, the radius of the photon sphere depends non-trivially on the nonlinearity parameter $\gamma$, leading to potentially observable effects in strong gravitational lensing scenarios \cite{liu2019probing}.

One of the most intriguing phenomena in black hole physics is the Aschenbach effect, a relativistic effect first identified in rapidly rotating Kerr black holes \cite{stuchlik2005aschenbach,khodagholizadeh2020aschenbach}. This effect manifests as a non-monotonic behavior of the angular velocity of particles in circular orbits as a function of radius, contradicting the standard Keplerian expectation that angular velocity should decrease monotonically with increasing radius \cite{stuchlik2005aschenbach,khodagholizadeh2020aschenbach}. Remarkably, our analysis reveals that this effect can also emerge in spherically symmetric EPYM black holes, despite the absence of rotation. This finding suggests that the nonlinearity of the YM field can mimic certain aspects of rotational effects in spacetime, creating regions where the effective gravitational force exhibits complex radial dependencies \cite{harikumar2024yang,jokela2019memory}.

In this work, we aim to bridge these various perspectives by systematically analyzing the EPYM black hole solution and its physical implications. We first derive the exact metric function for the EPYM AdS black hole and examine its horizon structure and singularities. We then investigate the thermodynamic properties, including temperature, entropy, and phase transitions, with particular attention to how the nonlinearity parameter $\gamma$ affects critical behavior. Next, we study the quantum tunneling of $W^+$ bosons from the EPYM black hole using the WKB approximation, deriving the tunneling probability and the corresponding Hawking temperature. This semiclassical approach allows us to verify the consistency of the quantum and thermodynamic descriptions while revealing quantum corrections induced by the nonlinear field \cite{soroushfar2024exploring,ali2024first,pourhassan2025thermal,pourhassan2020pv}. A significant portion of our analysis is devoted to understanding the effective potential governing particle motion and field propagation around EPYM black holes. By solving the Klein-Gordon equation in this context, we derive the effective potential and analyze its behavior for different values of the nonlinearity parameter $\gamma$. This analysis reveals how the nonlinear YM field modifies the stability regions and radiation characteristics compared to standard black hole solutions \cite{rincon2024quasinormal}. We also extract the effective force from this potential, showing how it transitions from attractive to repulsive regimes under certain conditions, with important implications for particle confinement and accretion processes. Our investigation also extends to photon orbits and null circular geodesics, which determine the optical appearance and shadow of the black hole. By solving the geodesic equations for null particles, we identify the conditions for circular orbits and analyze how their properties depend on the black hole parameters. We also establish a connection between these null geodesics and the orbits with extreme orbital periods, demonstrating that both satisfy the same mathematical condition despite arising from different physical considerations. Perhaps the most surprising result of our study is the identification of the Aschenbach effect in spherically symmetric EPYM black holes. Through a detailed analysis of the angular velocity profile for timelike circular orbits, we establish the conditions under which this effect emerges and relate it to the stability properties of photon spheres. This finding has profound implications for our understanding of relativistic orbital dynamics and suggests that nonlinear electromagnetic fields can induce effects previously thought to require spacetime rotation \cite{wei2024aschenbach}.

The paper is organized as follows. Section \ref{sec2} presents a comprehensive review of EPYM black hole solutions in AdS backgrounds, focusing on their metric structure and basic thermodynamic properties. Section \ref{sec3} examines the quantum tunneling of $W^+$ bosons in the EPYM black hole background, establishing the formalism and deriving expressions for the tunneling probability and Hawking temperature. Section \ref{sec4} analyzes the effective potential governing field propagation and particle motion, with emphasis on its role in radiation emission and its dependence on the nonlinearity parameter. Section \ref{sec5} focuses on photon orbits and null circular geodesics, exploring their properties and their connection to the black hole's optical characteristics. Section \ref{sec6} investigates the Aschenbach effect in EPYM black holes, deriving the conditions for its occurrence and discussing its physical significance. Finally, Section \ref{sec7} presents our conclusions and discusses potential directions for future research in this area.

\section{Comprehensive Review of Nonlinear Yang-Mills AdS Black Holes: Solutions and Thermodynamic Insights}\label{sec2}

Nonlinear electrodynamics has emerged as a significant area of research in gravitational physics, offering valuable insights into the behavior of spacetime under extreme conditions. The EPYM theory represents an important extension to standard Einstein-Maxwell theory by incorporating nonlinear YM fields characterized by a power parameter $\gamma$. This section provides a comprehensive review of EPYM black hole solutions in AdS backgrounds, focusing on their metric structure, horizon properties, and thermodynamic behavior.

The action describing four-dimensional EPYM gravity with a cosmological constant $\Lambda$ is given by Eq. \eqref{izaction} \cite{mazharimousavi2009lovelock}. This nonlinearity parameter $\gamma$ plays a crucial role in modifying the spacetime geometry and, consequently, the physical properties of the resulting black hole solutions \cite{du2023nonlinearity}. The field strength tensor for the YM field is expressed as:

\begin{equation}
F^{(a)}_{\mu\nu} = \partial_\mu A^{(a)}_{\nu} - \partial_\nu A^{(a)}_{\mu} + \frac{1}{2\zeta} C^{(a)}_{(b)(c)} A^b_{\mu} A^c_{\nu},
\end{equation}

where $C^{(a)}_{(b)(c)}$ are the structure constants of the SU(2) gauge group, reflecting the non-Abelian nature of the YM field. This non-Abelian character, combined with the nonlinear power term, introduces rich and complex behavior not observed in linear electromagnetic theories.

The corresponding black hole metric takes the standard spherically symmetric form:

\begin{equation} \label{izgm}
ds^2 = -f(r) dt^2 + f^{-1}(r) dr^2 + r^2 d\Omega^2_2,
\end{equation}

where $d\Omega^2_2$ represents the metric of a unit 2-sphere, and the function $f(r)$ is given by \cite{yerra2019heat}:

\begin{equation} \label{izm} 
f(r) = 1 - \frac{2M}{r} - \frac{\Lambda}{3} r^2 + \frac{(2q^2)^\gamma}{2(4\gamma - 3)r^{4\gamma - 2}}.
\end{equation}

This solution is valid for $\gamma \neq 3/4$, and the YM power term satisfies the weak energy condition (WEC) for $\gamma > 0$ \cite{yerra2019heat}. The parameter $q$ represents the YM charge, which measures the strength of the nonlinear electromagnetic field. The dimensional parameter $\gamma$ controls the degree of nonlinearity, with $\gamma = 1$ corresponding to the standard Yang-Mills case.

The event horizon of the EPYM black hole, denoted by $r_h$, is determined by solving the equation $f(r_h) = 0$. Unlike Schwarzschild or Reissner-Nordström black holes, the horizon structure of EPYM black holes exhibits more complex behavior due to the nonlinear term, potentially allowing for multiple horizons depending on the values of $M$, $q$, $\Lambda$, and $\gamma$ \cite{corda2011inflation}. The largest root of $f(r) = 0$ corresponds to the event horizon, which separates the interior region from the exterior spacetime accessible to distant observers.

By computing the expression $f(r_h) = 0$, we can express the mass parameter $M$ in terms of the horizon radius and other parameters:

\begin{equation}
M = \frac{r_h}{2} - \frac{\Lambda r_h^3}{6} + \frac{(2q^2)^\gamma}{2(4\gamma - 3)r_h^{4\gamma - 3}}.
\end{equation}

This relation proves essential for analyzing the thermodynamic properties of the black hole, as it connects the gravitational mass to the horizon geometry and field parameters \cite{bekenstein1973black,hawking1975particle}.

\section{Quantum Tunneling of Bosons in EPYM Black Hole Background}\label{sec3}

The quantum tunneling formalism represents one of the most elegant approaches to understanding Hawking radiation from black holes. This section explores the quantum tunneling of massive vector bosons, specifically the $W^+$ boson, in the background of Einstein-Power-Yang-Mills (EPYM) black holes. This investigation not only provides insights into the quantum nature of black hole radiation but also reveals how the nonlinear Yang-Mills (YM) parameter $\gamma$ influences the emission process \cite{du2023nonlinearity}.

The quantum tunneling approach, developed by Parikh and Wilczek and later extended to various particle species, offers a semiclassical picture of particle creation near black hole horizons \cite{parikh2000hawking}. Unlike Hawking's original derivation, which relied on quantum field theory in curved spacetime, the tunneling method visualizes radiation as a tunneling process through the classically forbidden region at the event horizon. This approach has proven particularly valuable for studying radiation from modified gravity black holes, where the spacetime geometry deviates from those in general relativity \cite{shi2023quantum,ren2023hawking,akhmedov2006hawking,angheben2005hawking}.

In the context of EPYM black holes, mass and charge are not concentrated at a single point but rather spread out due to the nonlinear electromagnetic field. To analyze the tunneling process of vector bosons, we need to consider the appropriate field equations in this curved spacetime background \cite{jusufi2017quantum,gecim2017massive}.

The Lagrangian that governs $W$ bosons in an electromagnetic field is given by \cite{li2015massive}:

\begin{align}
L&= -\frac{1}{2} ( D_{\mu}^+ W_{\nu}^+ - D_{\nu}^+ W_{\mu}^+ )( D^{-\mu} W^{-\nu} - D^{-\nu} W^{-\mu} )\notag \\
&+ m_W^2 W_{\mu}^+ W^{-\mu} - i e F_{\mu\nu} W_{\mu}^+ W^{-\nu}, 
\end{align}

where $D_{\mu}^{\pm} = \nabla_{\mu} \pm i e A_{\mu}$ is the covariant derivative and $A_{\mu} = (A_0, 0, 0, 0)$ is the electromagnetic potential of the black hole. The equation of motion for the $W$ boson field is:
\begin{align}
&\frac{1}{\sqrt{-g}} \partial_{\mu} \left[ \sqrt{-g} \left( D_{\nu}^{\pm} W_{\mu}^{\pm} - D_{\mu}^{\pm} W_{\nu}^{\pm} \right) \right] 
\notag \\
& \pm ie A_{\mu} \left( D_{\nu}^{\pm} W_{\mu}^{\pm} - D_{\mu}^{\pm} W_{\nu}^{\pm} \right) \notag \\
&+ m_W^2 W_{\nu}^{\pm} \pm i e F^{\nu\mu} W_{\mu}^{\pm} = 0.
\end{align}

In this equation, $F^{\mu\nu} = \nabla^\mu A^\nu - \nabla^\nu A^\mu$ represents the electromagnetic field tensor. Our analysis focuses specifically on the tunneling of the $W^+$ boson, though similar principles apply to other vector bosons as well \cite{ovgun2016massive}.

To solve these equations semiclassically, we apply the WKB (Wentzel-Kramers-Brillouin) approximation, which is valid when the particle's de Broglie wavelength is much smaller than the characteristic curvature radius of the spacetime. Under this approximation, we express the $W^+$ boson field as \cite{ovgun2016massive}:

\begin{equation}
W_{\mu}^+(t, r, \theta, \varphi) = b_{\mu}(t, r, \theta, \varphi) \exp \left( \frac{i}{\hbar} I(t, r, \theta, \varphi) \right), 
\end{equation}

where $I(t, r, \theta, \varphi)$ is the action and $b_{\mu}(t, r, \theta, \varphi)$ represents the amplitude. In the semiclassical limit ($\hbar \to 0$), we can neglect the higher-order terms in $\hbar$, keeping only the leading order. This approximation leads to a system of equations for the components of the vector field.

Substituting the WKB ansatz into the field equations and keeping only leading-order terms in $\hbar$, we obtain the following system of equations:

\begin{align}
\,& b_0 \left( -(\partial_1 I)^2 - \frac{(\partial_2 I)^2}{r^2 f(r)}  
- \frac{(\partial_3 I)^2}{r^2 f(r) \sin^2\theta} - \frac{m^2}{f(r)} \right)  \nonumber \\
&+ b_1 (\partial_1 I) (e A_0 + \partial_0 I)  \nonumber \\
&+ b_2 \left( \frac{(\partial_2 I)}{r^2 f(r)} (e A_0 + \partial_0 I) \right)  \nonumber \\
&+ b_3 \left( \frac{(\partial_3 I)}{r^2 f(r) \sin^2\theta} (e A_0 + \partial_0 I) \right)=0,
\end{align}

\begin{align}
\,& b_0 (-\partial_1 I (e A_0 + \partial_0 I))  \nonumber \\
&+ b_1 \bigg(  \frac{-f(r)(\partial_2 I)^2}{r^2}  
-  \frac{f(r)(\partial_3 I)^2}{r^2 \sin^2\theta} \notag \\ 
&+ (e A_0 + \partial_0 I)^2 - m^2 f(r) \bigg)  \nonumber \\
&+ b_2 \left( f(r) \frac{\partial_1 S \partial_2 I}{r^2} \right)  \nonumber \\
&+ b_3 \left( f(r) \frac{\partial_1 I \partial_3 I}{r^2 \sin^2\theta} \right)=0,
\end{align}

\begin{align}
\,& b_0 \left( -\frac{\partial_2 I}{f(r)} (e A_0 + \partial_0 I) \right)  \nonumber \\
&+ b_1 (f(r) \partial_2 I \partial_1 I)  \nonumber \\
&+ b_2 \left( -f(r) (\partial_1 I)^2 - \frac{(\partial_3 I)^2}{r^2 \sin^2\theta}  
+ \frac{(e A_0 + \partial_0 I)^2}{f(r)} - m^2 \right)  \nonumber \\
&+ b_3 \left( \frac{\partial_2 I \partial_3 I}{r^2 \sin^2\theta} \right)=0,
\end{align}

\begin{align}
\,& b_0 \left( -\frac{\partial_3 I}{f(r)} (e A_0 + \partial_0 I) \right)  \nonumber \\
&+ b_1 (f(r) \partial_3 I \partial_1 I)  \nonumber \\
&+ b_3 \left( -f(r) (\partial_1 I)^2 - \frac{(\partial_2 I)^2}{r^2}  
+ \frac{(e A_0 + \partial_0 I)^2}{f(r)} - m^2 \right)  \nonumber \\
&+ b_2 \left( \frac{\partial_2 I \partial_3 I}{r^2} \right)=0.
\end{align}

These coupled equations represent the behavior of the vector boson field in the vicinity of the EPYM black hole. For a non-trivial solution to exist, the determinant of the coefficient matrix must vanish, which leads to constraints on the possible values of the action $I(t, r, \theta, \varphi)$.

Considering the spherical symmetry of the spacetime, we can form a coefficient matrix $\xi$ whose elements are:

\begin{align}
\xi_{11} &= -(\partial_1 I)^2 - \frac{(\partial_2 I)^2}{r^2 f(r)} 
- \frac{(\partial_3 I)^2}{r^2 f(r) \sin^2\theta} - \frac{m^2}{f(r)}, \\
\xi_{12} &= (\partial_1 I) (e A_0 + \partial_0 I), \\
\xi_{13} &= \frac{(\partial_2 I)}{r^2 f(r)} (e A_0 + \partial_0 I), \\
\xi_{14} &= \frac{(\partial_3 I)}{r^2 f(r) \sin^2\theta} (e A_0 + \partial_0 I), \\
\xi_{21} &= -(\partial_1 I) (e A_0 + \partial_0 I), \\
\xi_{22} &= -f(r) \frac{(\partial_2 I)^2}{r^2} - f(r) \frac{(\partial_3 I)^2}{r^2 \sin^2\theta}  
+ (e A_0 + \partial_0 I)^2 - m^2 f(r), \\
\xi_{23} &= f(r) \frac{\partial_1 I \partial_2 I}{r^2}, \\
\xi_{24} &= f(r) \frac{\partial_1 I \partial_3 I}{r^2 \sin^2\theta}, \\
\xi_{31} &= -\frac{\partial_2 I}{f(r)} (e A_0 + \partial_0 I), \\
\xi_{32} &= f(r) \partial_2 I \partial_1 I, \\
\xi_{33} &= -f(r) (\partial_1 I)^2 - \frac{(\partial_3 I)^2}{r^2 \sin^2\theta}  
+ \frac{(e A_0 + \partial_0 I)^2}{f(r)} - m^2, \\
\xi_{34} &= \frac{\partial_2 I \partial_3 I}{r^2 \sin^2\theta}, \\
\xi_{41} &= -\frac{\partial_3 I}{f(r)} (e A_0 + \partial_0 I), \\
\xi_{42} &= f(r) \partial_3 I \partial_1 I, \\
\xi_{43} &= \frac{\partial_2 I \partial_3 I}{r^2}, \\
\xi_{44} &= -f(r) (\partial_1 I)^2 - \frac{(\partial_2 I)^2}{r^2}  
+ \frac{(e A_0 + \partial_0 I)^2}{f(r)} - m^2.
\end{align}

To solve for the action, we exploit the symmetries of spacetime. Given the static and spherically symmetric nature of the EPYM black hole, we can adopt the following separation ansatz for the action \cite{Sakalli:2015jaa,Sakalli:2012zy,Sakalli:2023pgn,Vanzo:2011wq,Kubiznak:2016qmn,Jusufi:2017trn}:

\begin{equation}
S = -E t + W(r) + j \varphi + H(\theta) + b,
\end{equation}

where $E$ represents the energy of the emitted particle, $j$ is the angular momentum, and $b$ is a constant. Substituting this ansatz into the determinant equation and solving for the radial function $W(r)$, we obtain:

\begin{align}
 W_{\pm} &= \pm \int dr \frac{1}{f(r)} \bigg[E^2 - 2EeA_0 + e^2A_0^2 \notag \\
 &- f(r) \left(m^2+\frac{(\partial_2 H)^2}{r^2}\right) \bigg]^{\frac{1}{2}}. 
\end{align}
This integral describes the radial motion of the $W^+$ boson as it tunnels through the event horizon. The signs $+$ and $-$ correspond to the outgoing and ingoing solutions, respectively. The tunneling probability is related to the imaginary part of this action\cite{angheben2005hawking,Sakalli:2022xrb}.

To evaluate this integral near the horizon, we expand the metric function $f(r)$ in the vicinity of $r_h$:

\begin{equation}
f(r_h) \approx f'(r_h)(r - r_h), 
\end{equation}

where the prime denotes differentiation with respect to $r$. Using this approximation, we can perform the integration across the pole at $r = r_h$, which yields:

\begin{equation}
W_{\pm} = \pm i \pi \frac{\sqrt{E^2 - 2EeA_0 + e^2A_0^2}}{f'|_{r_h}}. 
\end{equation}

From the standard quantum tunneling formalism, the tunneling probability is given by \cite{ovgun2016massive}:

\begin{equation} 
\Gamma = e^{-4 \text{Im} W_+} = e^{- \frac{E_{balance}}{T_H}}, 
\end{equation}

where $E_{balance} = E - eA_0$ represents the effective energy of the tunneling particle, accounting for electromagnetic interactions. Comparing this expression with the Boltzmann factor for thermal radiation, we can identify the Hawking temperature as:

\begin{equation}
T_H = \frac{1}{4\pi} \left. \frac{df(r)}{dr} \right|_{r_h}.
\end{equation}

By substituting the explicit form of $f(r)$ for the EPYM black hole, we obtain the expression for the Hawking temperature:

\begin{equation}
T_H= \frac{M}{2\pi r_h^2} - \frac{\Lambda}{6\pi} r_h - \frac{(2q^2)^\gamma (4\gamma - 2)}{8\pi (4\gamma - 3) r_h^{4\gamma - 1}}.
\end{equation}

This expression reveals how the nonlinear YM parameter $\gamma$ affects the black hole's thermal radiation. For different values of $\gamma$, the temperature profile exhibits distinct behaviors, especially at small horizon radii where nonlinear effects become dominant.

\begin{figure}[h]
    \centering
    \includegraphics[width=0.5\textwidth]{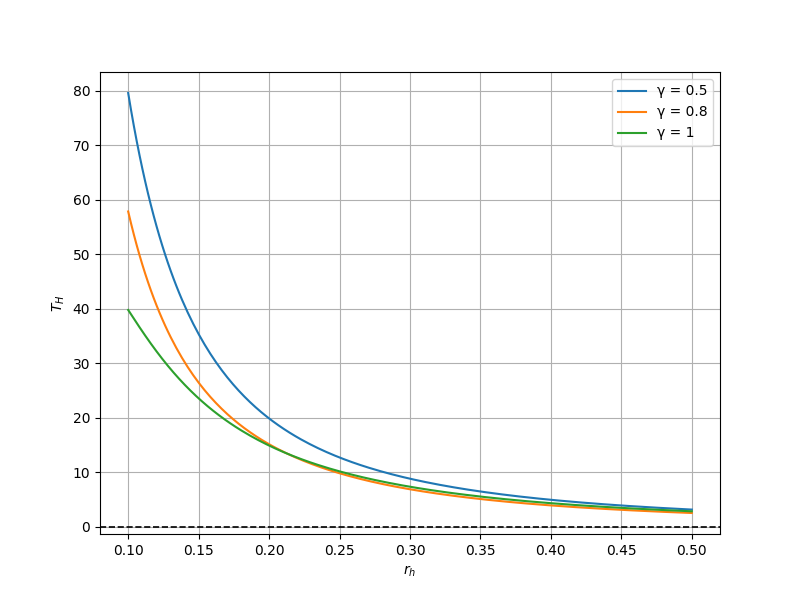}
    \caption{Plot of Hawking temperature $T_H$ as a function of the event horizon radius $r_h$ for different values of $\gamma$. The parameters are set to $M = 5$, $\Lambda = 0.1$, and $q = 0.5$. The dashed line represents $T_H = 0$, indicating the extremal black hole condition.}
    \label{hawking_graph}
\end{figure}

The variation of Hawking temperature with horizon radius for different values of $\gamma$ reveals important insights into the thermodynamic stability of EPYM black holes. As shown in Figure \ref{hawking_graph}, the temperature is particularly high for small values of $r_h$ and decreases rapidly as $r_h$ increases. The nonlinearity parameter $\gamma$ significantly affects the temperature profile, with smaller values of $\gamma$ leading to more pronounced temperature increases at small radii. This behavior has important implications for the final stages of black hole evaporation, suggesting that nonlinear electromagnetic effects could substantially modify the standard picture of black hole thermodynamics.

The tunneling approach provides a physically intuitive picture of Hawking radiation from EPYM black holes and allows us to incorporate the effects of nonlinear electrodynamics on the emission process. This analysis not only confirms the thermodynamic temperature derived through other methods but also offers insights into how massive vector bosons are emitted from these black holes. The dependence of the tunneling probability on the nonlinearity parameter $\gamma$ highlights the influence of nonlinear electrodynamics on quantum processes in strong gravitational fields.

\section{Analysis of Effective Potential and Its Role in Radiation Emission}\label{sec4}

The effective potential governing particle motion in black hole spacetimes provides crucial insights into various astrophysical phenomena, including radiation emission processes, particle confinement, and the stability of orbits. For EPYM black holes, the effective potential exhibits unique features due to the nonlinear YM field, which significantly alters the spacetime geometry compared to standard solutions in general relativity. This section presents a detailed analysis of the effective potential for scalar field propagation around EPYM black holes and examines its implications for radiation processes.

The effective potential approach has proven invaluable in understanding black hole physics across various contexts. In classical general relativity, it provides a framework for analyzing the stability of particle orbits and the behavior of test fields in black hole backgrounds \cite{s2024effective,igata2020stable}. In semiclassical gravity, the effective potential governs the greybody factors that modify the black body spectrum of Hawking radiation, accounting for the backscattering of emitted particles by the spacetime curvature. The unique features of the effective potential in modified gravity theories like EPYM can lead to distinctive observational signatures that potentially differentiate these black holes from their counterparts in general relativity.

The nonlinear nature of the YM field in EPYM theory introduces additional complexity to the effective potential, particularly through the parameter $\gamma$ that controls the degree of nonlinearity. This parameter significantly influences the shape and behavior of the potential, especially at small radial distances where the nonlinear electromagnetic effects become dominant \cite{chakhchi2022shadows}. Understanding these modifications is essential for predicting the radiation spectrum, absorption cross-sections, and quasinormal modes of EPYM black holes, which could potentially be detected by current and future gravitational wave observatories.

In this section, we obtain the effective potential by analyzing the propagation of scalar fields in the background of the static and spherically symmetric EPYM black hole \eqref{izgm} defined by metric function \eqref{izm}. While this approach focuses on scalar fields for simplicity, the qualitative features extend to higher-spin fields as well, with appropriate modifications to account for spin-curvature coupling. The resulting effective potential encapsulates the essential physics of wave propagation in this curved spacetime and provides a foundation for understanding more complex radiation processes.

To analyze the propagation of a scalar field in the EPYM black hole background, we begin with the Klein-Gordon equation for a minimally coupled scalar field $\Xi$:

\begin{equation}
\partial_\mu [\sqrt{-g} g^{\mu\nu} \partial_\nu \Xi] = 0.
\end{equation}

This equation describes the behavior of a massless scalar field in curved spacetime, assuming minimal coupling between the field and gravity \cite{biswas2022non,buchdahl1959reciprocal}. For the spherically symmetric spacetime under consideration, we can separate the variables using the decomposition method:

\begin{equation}
\Xi = e^{-i\omega t} R_{\omega l m}(r) Y_l^m(\theta, \phi).
\end{equation}

Here, $\omega$ represents the frequency of the field, $R_{\omega l m}(r)$ is the radial function, and $Y_l^m(\theta, \phi)$ are the spherical harmonics that capture the angular dependence of the field. Substituting this decomposition into the Klein-Gordon equation and using the metric components of the EPYM black hole, we obtain separate equations for the radial and angular parts:

\begin{equation}
\frac{1}{r^2} \frac{d}{dr} \left( r^2 f \frac{dR_{\omega l m}}{dr} \right) + \left( \frac{\omega^2}{f} - \frac{\lambda_l}{r^2} \right) R_{\omega l m} = 0,
\end{equation}

\begin{equation}
\frac{1}{\sin \theta} \frac{\partial}{\partial \theta} \left( \sin \theta \frac{\partial Y_l^m}{\partial \theta} \right) + \frac{1}{\sin^2 \theta} \frac{\partial^2 Y_l^m}{\partial \phi^2} + \lambda_l Y_l^m = 0.
\end{equation}

The angular equation is the standard eigenvalue equation for spherical harmonics, with eigenvalue $\lambda_l$ related to the angular momentum quantum number $l$. For a scalar field in spherically symmetric spacetime, $\lambda_l = l(l+1)$. In more general contexts, particularly for rotating black holes, the separation constant may have a more complex form, as given by the power series expansion \cite{berti2006eigenvalues}:

\begin{equation}
\lambda_l = \sum_{k=0}^{\infty} (a \omega)^k F_{l,k}, \quad \lambda_l = l(l+1) 
\end{equation}

where $a$ is the spin parameter of the black hole and $F_{l,k}$ are coefficients dependent on the specific problem. The angular momentum satisfies conditions $l \geq |m|$ and $\frac{l - |m|}{2} \in (0, \mathbb{Z})$.

To derive the effective potential that governs the radial propagation of the field, we introduce the tortoise coordinate $v^\star$ defined by the relation:

\begin{equation}
\frac{dv^\star}{dr} = \frac{1}{f}, \quad \frac{d}{dv^\star} = f \frac{d}{dr}, \quad \frac{d^2}{dv^{\star 2}} = f \left( \frac{d^2}{dr^2} + \frac{df}{dr} \frac{d}{dr} \right).
\end{equation}

This coordinate transformation is standard in black hole physics and has the advantage of mapping the event horizon to $v^\star \to -\infty$ while asymptotic infinity corresponds to $v^\star \to \infty$. This transformation simplifies the wave equation and provides a clear physical interpretation in terms of wave propagation. Additionally, we rescale the radial function as:

\begin{equation}
R_{\omega l m}(r) = \frac{T_{\omega l m}(r)}{r},
\end{equation}

which removes the first-derivative term in the radial equation. After these transformations, the radial equation becomes:

\begin{equation}
\left( \frac{d^2}{dv^{\star 2}} - V_{\text{eff}} \right) T_{\omega l m} = 0,
\end{equation}

where the effective potential $V_{\text{eff}}$ is given by:

\begin{equation}
V_{\text{eff}} = f(r) \left( \frac{1}{r} \frac{d f(r)}{dr} - \omega^2 + \frac{l(l+1)}{r^2} \right)
\end{equation}

Substituting the specific form of $f(r)$ for the EPYM black hole, we obtain:

\begin{align}
V_{\text{eff}} &= \left( 1 - \frac{2M}{r} - \frac{\Lambda}{3} r^2 + \frac{(2q^2)^\gamma}{2(4\gamma - 3)r^{4\gamma - 2}} \right) \times \notag \\
& \quad \left( \frac{2M}{r^3} - \frac{2\Lambda}{3} - \frac{(2q^2)^\gamma(4\gamma - 2)}{2(4\gamma - 3)r^{4\gamma}} - \omega^2 + \frac{l(l+1)}{r^2} \right)
\end{align}

This expression reveals how the effective potential depends on the black hole parameters ($M$, $q$, $\Lambda$), the nonlinearity parameter $\gamma$, the field frequency $\omega$, and the angular momentum quantum number $l$. The behavior of this potential determines various physical properties of the black hole, including its stability, radiation spectrum, and quasinormal modes.

Figure \ref{v_eff} illustrates the variation of the effective potential $V_{\text{eff}}$ with respect to the radial coordinate $r$ for different values of the nonlinearity parameter $\gamma$. The graph reveals several important features of the potential profile:

\begin{figure}[h]
    \centering
    \includegraphics[width=0.5\textwidth]{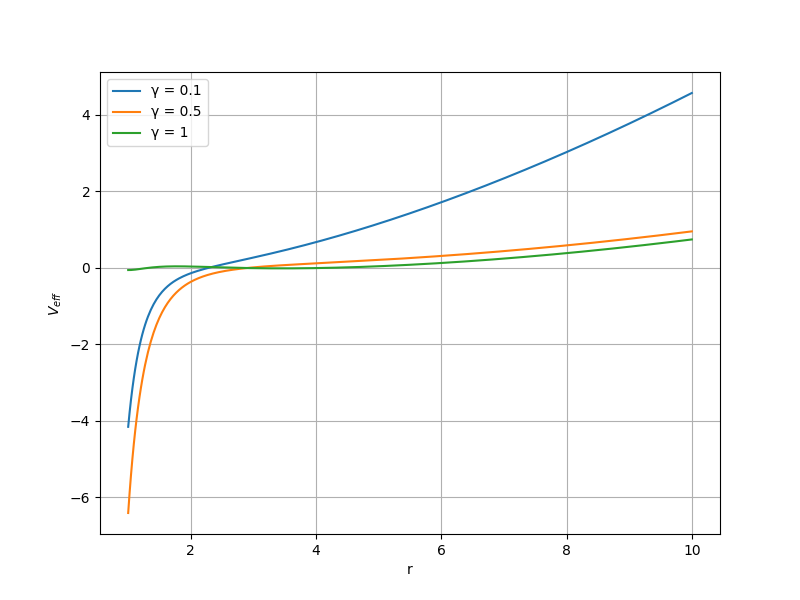}
    \caption{Variation of the effective potential $V_{\text{eff}}$ as a function of $r$ for different values of $\gamma$ ($\gamma = 0.1, 0.5, 1$). The parameters used are $M = 1$, $q = 1$, $\Lambda = 0.1$, $\omega = 0.5$, and $l = 1$. The behavior of $V_{\text{eff}}$ changes significantly for small $r$ values, especially for lower $\gamma$, while converging for larger $r$.}
    \label{v_eff}
\end{figure}

For small values of $r$ ($r < 3$), the effective potential shows a steep decrease to negative values, with this decline more pronounced for smaller values of $\gamma$ ($\gamma = 0.1, 0.5$). This behavior indicates the strong influence of nonlinear electromagnetic corrections in regions close to the black hole. In the intermediate region ($3 \leq r \leq 6$), the potential approaches zero, while for large values of $r$ ($r > 6$), all curves converge to positive values as the effects of the cosmological constant become dominant.

The shape of the effective potential has important implications for radiation processes. Regions where the potential forms a barrier (positive values) tend to reflect incoming waves, while regions with negative potential values allow for transmission \cite{iyer1987black}. For EPYM black holes, the potential barrier is modified by the nonlinearity parameter $\gamma$, which affects both the height and width of the barrier. Lower values of $\gamma$ create a deeper negative well near the horizon, potentially enhancing the emission of low-frequency radiation, while also forming a higher barrier at intermediate distances that could suppress high-frequency emission \cite{sheikhahmadi2023astrophysical}.

\subsection{Effective Force}

Beyond the effective potential, the effective force experienced by test particles provides another important perspective on the gravitational dynamics around EPYM black holes. The effective force serves as an indicator of whether a test particle in the gravitational field is attracted toward or repelled from the central mass. Mathematically, the effective force experienced by a particle in a gravitational field is given by:

\begin{equation}
F = -\frac{1}{2} \frac{\partial V_{\text{eff}}}{\partial r}.
\end{equation}

For the EPYM black hole metric considered in this study, the effective force takes the form:

{\small
\begin{align}
F_{eff} &=  \left( \frac{1}{2} - \frac{M}{r} - \frac{\Lambda}{6} r^2 
+ \frac{(2q^2)^\gamma}{4(4\gamma - 3)r^{4\gamma - 2}} \right)  
\times \Bigg( \frac{6M}{r^4} - \frac{2l(l+1)}{r^3} \notag \\
&\quad - \frac{(2q^2)^\gamma(4\gamma - 2)(4\gamma)}
{2(4\gamma - 3)r^{4\gamma +1}} \Bigg)  \notag \\
&\quad + \Bigg( -\frac{M}{r^3} + \frac{\Lambda}{3}  
+ \frac{(2q^2)^\gamma(4\gamma - 2)}{4(4\gamma - 3)r^{4\gamma}}  \notag \\
&\quad + \frac{\omega^2}{2} - \frac{l(l+1)}{2r^2} \Bigg) 
\times \Bigg( \frac{2M}{r^2} - \frac{2\Lambda}{3} r  
- \frac{(2q^2)^\gamma (4\gamma - 2)}{2(4\gamma - 3) r^{4\gamma - 1}} \Bigg).
\end{align}
}

This complex expression reflects the interplay between various factors affecting particle dynamics in the EPYM black hole spacetime. The effective force depends not only on the black hole parameters ($M$, $q$, $\Lambda$) and the nonlinearity parameter $\gamma$ but also on the frequency parameter $\omega$ and the angular momentum quantum number $l$ of the test particle. The radial dependence of the force reveals how particles at different distances from the black hole experience different gravitational effects.

Figure \ref{f_eff} illustrates the variation of the effective force $F_{\text{eff}}$ with respect to the radial coordinate $r$ for different values of the nonlinearity parameter $\gamma$:

\begin{figure}[h]
    \centering
    \includegraphics[width=0.5\textwidth]{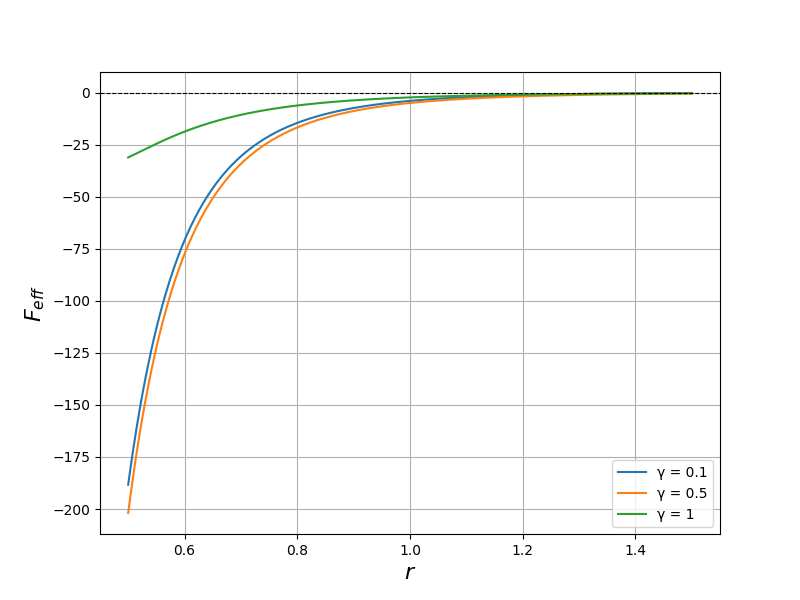}
    \caption{The variation of the effective force $F_{\text{eff}}$ with respect to the radial coordinate $r$ for different values of $\gamma$. The parameters are chosen as $M = 1$, $q = 0.5$, $\Lambda = 0.1$, $l = 1$, and $\omega = 0.5$. The results indicate that for smaller values of $\gamma$, the attractive force is stronger near the center, while for larger $\gamma$, the force weakens and transitions to a repulsive regime at larger $r$.}
    \label{f_eff}
\end{figure}

The graph reveals several important features of the force profile. For small values of $r$, the effective force is negative, indicating an attractive force that pulls the test particles toward the central mass. For smaller values of $\gamma$ ($\gamma = 0.1$ and $\gamma = 0.5$), the magnitude of this attractive force is significantly higher, suggesting a stronger gravitational pull in regions close to the black hole. In contrast, for larger values of $\gamma$ ($\gamma = 1$), the force has a smaller magnitude and transitions to positive values (repulsive force) at larger radii.

These results demonstrate how the nonlinearity parameter $\gamma$ fundamentally alters the gravitational dynamics around EPYM black holes. Lower values of $\gamma$ enhance the attractive force, making it more difficult for particles to escape from regions close to the black hole. This has important implications for accretion processes, as the enhanced gravitational pull could lead to higher accretion rates and more efficient energy extraction. Conversely, higher values of $\gamma$ weaken central attraction and introduce repulsive effects at larger distances, potentially creating stable orbits where standard general relativity would predict instability.

The effective potential and force analyses provide valuable insights into the radiation emission processes from EPYM black holes. The modified potential barrier affects the transmission and reflection coefficients for waves propagating in this spacetime, directly influencing the spectrum of emitted radiation \cite{guo2020scalar}. The deeper negative well near the horizon for lower $\gamma$ values suggests enhanced emission of low-frequency radiation, while the higher barrier at intermediate distances could suppress high-frequency components, leading to a modified black-body spectrum compared to standard Schwarzschild black holes \cite{volkov1989non}. Additionally, the effective force profile affects the dynamics of charged particles around the black hole, which can generate secondary radiation through processes like synchrotron emission and bremsstrahlung. The stronger attractive force for lower $\gamma$ values could accelerate charged particles to higher energies, potentially leading to more energetic radiation signatures. These distinctive features in the radiation spectrum could provide observational tests for the existence of nonlinear electromagnetic effects in astrophysical black holes.
\section{Photon Orbits and Null Circular Geodesics in EPYM Black Holes}\label{sec5}

The study of null geodesics, particularly circular-photon orbits, provides crucial insights into the optical properties and causal structure of black hole spacetimes. For EPYM black holes, these geodesics exhibit distinctive features due to the nonlinear YM field, which significantly alters the spacetime geometry compared to solutions in general relativity. This section presents a detailed analysis of null circular geodesics in EPYM black holes and examines their implications for observable phenomena such as black hole shadows and gravitational lensing.

Photon orbits play a fundamental role in determining the optical appearance of black holes to distant observers \cite{cunha2017fundamental}. The photon sphere, a region where light can travel in unstable circular orbits, forms the boundary of the black hole shadow and governs strong gravitational lensing effects \cite{gralla2019black}. In standard general relativity, the photon sphere of a Schwarzschild black hole is located at $r = 3M$, while for charged and rotating black holes, its position depends on the charge and spin parameters. For EPYM black holes, the nonlinear YM field introduces additional complexity, with the photon sphere location depending non-trivially on the nonlinearity parameter $\gamma$.

In this section, we investigate the physical and mathematical properties of null circular geodesics in the background of EPYM black holes. Our analysis focuses primarily on null circular geodesics in the equatorial plane ($\theta = \pi/2$), which captures the essential features of the orbital dynamics of photons. Following the study \cite{cardoso2009geodesic}, we obtain the equation of a null circular geodesic. The Lagrangian that defines the geodesics is as follows:

\begin{equation}
2\mathcal{L} = -f(r) \dot{t}^2 + f^{-1}(r) \dot{r}^2 + r^2 \dot{\phi}^2.
\end{equation}

Here, a dot denotes the derivative with respect to an affine parameter along the geodesic. Since the Lagrangian does not depend on the coordinates $t$ and $\phi$, the corresponding conserved quantities are called $E$ (energy) and $L$ (angular momentum). The generalized momentum components derived from the Lagrangian are expressed as follows \cite{cardoso2009geodesic}:

\begin{equation}
p_t = g_{tt} \dot{t} = -f(r) \dot{t} = -E = \text{const},
\end{equation}

\begin{equation}
p_\phi = g_{\phi\phi} \dot{\phi} = r^2 \dot{\phi} = L = \text{const},
\end{equation}

\begin{equation}
p_r = g_{rr} \dot{r} = f^{-1}(r) \dot{r}.
\end{equation}

The Hamiltonian of the system is given by:

\begin{equation}
H = p_t \dot{t} + p_r \dot{r} + p_\phi \dot{\phi} - L
\end{equation}

and satisfies the condition:

\begin{equation}
2H = -E \dot{t} + L \dot{\phi} + g_{rr} \dot{r}^2 = \xi = \text{const}
\end{equation}

Here, $\xi = 0$ is taken for null geodesics, which corresponds to the case of massless particles such as photons.

From the conserved quantities, we obtain the following relations:

\begin{equation}
\dot{t} = \frac{E}{f(r)}, \quad \dot{\phi} = \frac{L}{r^2}.
\end{equation}

When these are substituted into the Hamiltonian constraint equation, we get:

\begin{equation}
\dot{r}^2 = f(r) \left( \frac{E^2}{f(r)} - \frac{L^2}{r^2} \right).
\end{equation}

For circular orbits, we require $\dot{r} = 0$ and $\ddot{r} = 0$, which means that the radial motion is constrained to a fixed radius. The first condition gives the following.

\begin{equation}
\frac{E^2}{f(r)} = \frac{L^2}{r^2}
\end{equation}

Taking the derivative of the radial equation and setting it to zero (the second condition), we obtain the following.

\begin{equation}
2f(r) - r f'(r) = 0.\label{null_circular}
\end{equation}

This equation determines the radius of null circular geodesics in the EPYM black hole spacetime. It's worth noting that these geodesics are associated with the orbits having the shortest orbital period around the central black hole. According to asymptotic observers, in order to minimize the orbital period for a given radius, one should move as close to the speed of light as possible. In this case, the orbital period is expressed as follows \cite{cardoso2009geodesic}:

\begin{equation}
T(r) = \frac{2\pi r}{\sqrt{f(r)}}.
\end{equation}

The circular motion with the shortest orbital period must satisfy the following condition:

\begin{equation}
T'(r) = 0.
\end{equation}

This condition gives the equation for the fastest circular orbit:

\begin{equation}
2f(r) - r f'(r) = 0.\label{fastest_circular}
\end{equation}

Substituting the specific form of $f(r)$ for the EPYM black hole, we obtain:

\begin{equation}
2 - \frac{6M}{r} + \frac{(2q^2)^\gamma}{(4\gamma - 3) r^{4\gamma - 2}} + \frac{(2q^2)^\gamma (4\gamma - 2)}{2(4\gamma - 3) r^{4\gamma - 2}} = 0
\end{equation}

An important observation here is that the equation for the null circular geodesic \eqref{null_circular} and the equation for the fastest circular orbit \eqref{fastest_circular} are identical. In other words, the equation for the extreme period circular radius and the null circular geodesic have the same roots. This mathematical coincidence has a deep physical meaning: null geodesics represent the limit of timelike geodesics as the particle velocity approaches the speed of light, and this limit also corresponds to the minimum possible orbital period \cite{mazharimousavi2024null}.

If we can demonstrate the existence of the fastest circular orbit, we will have proven that null circular geodesics also exist in the outer region of the EPYM black hole. Therefore, the function $T(r)$ must have a minimum at some finite radius $r = r_{\text{extrem}}$. At this point, equations \eqref{null_circular} and \eqref{fastest_circular} are satisfied, which shows that $r = r_{\text{extrem}}$ corresponds to the location of the null circular geodesics.

Table \ref{tab:rf_solutions} presents the numerical solutions for the radius of the fastest circular orbit $r_f$ (which coincides with the photon sphere radius) for different values of the charge parameter $q$ and the nonlinearity parameter $\gamma$, while keeping the black hole mass $M = 1$ and the cosmological constant $\Lambda = 1$ fixed.

\begin{table}[h]
    \centering
    \renewcommand{\arraystretch}{1.5}  
    \setlength{\tabcolsep}{15pt}       
    \begin{tabular}{|c|c|c|}
        \hline
        $q$ & $\gamma$ & $r_f$ \\
        \hline
        0.1 & 0.3 & 1.057 \\
        \hline
        0.1 & 0.5 & 1.076 \\
        \hline
        0.1 & 0.8 & 2.000 \\
        \hline
        0.1 & 1.0 & 2.000 \\
        \hline
        0.5 & 0.3 & 1.183 \\
        \hline
        0.5 & 0.5 & 1.547 \\
        \hline
        0.5 & 0.8 & 158.43 \\
        \hline
        0.5 & 1.0 & 2.000 \\
        \hline
        1.0 & 0.3 & 1.354 \\
        \hline
        1.0 & 0.5 & 3.414 \\
        \hline
        1.0 & 0.8 & 59608.80 \\
        \hline
        1.0 & 1.0 & 4.000 \\
        \hline
    \end{tabular}
    \caption{The table presents the solutions for $r_f$ with fixed $M = 1$ and $\Lambda = 1$, for different values of $q$ and $\gamma$.}
    \label{tab:rf_solutions}
\end{table}

The results in Table \ref{tab:rf_solutions} reveal several important features of photon orbits in EPYM black holes. For small values of $q$ and $\gamma$, the radius of the photon sphere remains close to the Schwarzschild-like value, around $r_f \approx 1.057 - 1.076$ for $q = 0.1$. This indicates that weak nonlinear electromagnetic effects do not significantly alter the photon orbit structure. However, as $q$ increases, the effects of nonlinear electrodynamics become more pronounced, leading to significantly higher values of $r_f$, particularly for $\gamma = 0.8$, where $r_f$ increases drastically (e.g., $r_f \approx 158.43$ for $q = 0.5$ and $r_f \approx 59608.80$ for $q = 1$).

This dramatic increase in the photon sphere radius suggests that higher values of $\gamma$ introduce strong repulsive effects that push the photon orbit far from the black hole. This behavior is particularly significant for observational implications, as it would lead to a much larger black hole shadow than predicted by general relativity, potentially providing a clear observational signature of nonlinear electromagnetic effects. Furthermore, for $\gamma = 1.0$, the values of $r_f$ remain relatively stable at $r_f = 2.000$ or $4.000$, indicating a potential threshold where the influence of the charge diminishes.

The existence and properties of photon orbits have important implications for various astrophysical phenomena. The photon sphere forms the boundary of the black hole shadow, determining its apparent size and shape as seen by distant observers. The significantly larger photon sphere radius for certain parameter values in EPYM black holes would result in a correspondingly larger shadow, potentially observable with current and future very long baseline interferometry (VLBI) facilities like the Event Horizon Telescope. Additionally, the photon sphere plays a crucial role in gravitational lensing phenomena, particularly in the formation of relativistic images \cite{virbhadra2002gravitational,virbhadra2000schwarzschild,claudel2001geometry}. These images arise when light rays from distant sources pass close to the black hole and undergo multiple loops around the photon sphere before reaching the observer. The properties of these relativistic images, including their magnification, separation, and time delays, depend sensitively on the photon sphere radius and could provide another observational test for nonlinear electromagnetic effects in black hole spacetimes.

\section{Aschenbach Effect in EPYM BH}\label{sec6}

The Aschenbach effect represents one of the most intriguing relativistic phenomena in black hole physics, characterized by a non-monotonic behavior of the angular velocity profile of orbiting particles. Initially discovered by Bernd Aschenbach in the context of rapidly rotating Kerr black holes \cite{aschenbach2007measurement}, this effect manifests itself as a surprising reversal in the expected monotonic decrease of angular velocity with increasing orbital radius. While traditionally associated with rotating black holes due to frame-dragging effects, recent studies have suggested that similar non-monotonic behaviors might emerge in certain non-rotating spacetimes under specific conditions. The presence of this effect in static, spherically symmetric EPYM black holes would represent a significant finding, suggesting that nonlinear electromagnetic fields can induce effects previously thought to require spacetime rotation.

The Aschenbach effect has important astrophysical implications, particularly for accretion dynamics and jet formation mechanisms around black holes \cite{aschenbach2004measuring,aschenbach2007measurement}. It may lead to distinctive observational signatures in the X-ray spectra of accreting black holes, including specific quasi-periodic oscillations (QPOs) that have been observed in both stellar-mass and supermassive black hole systems \cite{ingram2019review,caballero2013quasi,varniere2016possible}. Furthermore, the regions where angular velocity exhibits non-monotonic behavior could serve as potential sites for particle acceleration and energy extraction, influencing relativistic jet formation and high-energy phenomena around black holes.

In EPYM black holes, the nonlinear YM field introduces additional complexity to the spacetime geometry, potentially creating conditions where the Aschenbach effect could emerge even in the absence of rotation. The nonlinearity parameter $\gamma$ plays a crucial role in determining whether and how this effect manifests itself, with different values potentially leading to qualitatively different orbital velocity profiles \cite{du2023photon}. This section examines the necessary conditions for the Aschenbach effect to manifest in EPYM black holes and analyzes its physical implications.

The Aschenbach effect \cite{aschenbach2004measuring} refers to the non-monotonic behavior of the angular velocity of a timelike circular orbit (TCO) as a function of radial distance in a black hole spacetime. In the case of a zero-angular-momentum observer (Bardeen observer), this effect has been extensively studied for rapidly rotating black holes. However, its existence in static and spherically symmetric black holes remains an open question. In this section, we analyze the necessary conditions for the Aschenbach effect to manifest in such nonrotating black hole backgrounds. For a general static, spherically symmetric black hole described by the metric Eq. \eqref{izgm}, the angular velocity $\Omega_{CO}$ of a TCO as observed from a distant observer is given by the following:
   
\begin{equation}
\Omega_{CO} = \frac{\sqrt{f'(r)}}{\sqrt{2} r}. 
\end{equation}

This expression directly relates the angular velocity to the metric function and its derivative. For standard black holes in general relativity, such as the Schwarzschild solution, this angular velocity decreases monotonically with increasing radius, following a behavior similar to Keplerian orbits in Newtonian gravity. However, in the presence of nonlinear electromagnetic fields, as in EPYM black holes, this monotonic behavior can be disrupted.

To determine whether the Aschenbach effect is present, we examine the derivative of $\Omega_{CO}$ with respect to the radial coordinate.

\begin{equation}
\Omega'_{CO} = \frac{r f''(r) - f'(r)}{2\sqrt{2} r^3 \sqrt{f'(r)}}. 
\end{equation}

The Aschenbach effect occurs if there exists a radial region where $\Omega'_{CO} > 0$, indicating an increasing angular velocity with radius. This condition simplifies to:

\begin{equation}
r f''(r) - f'(r) > 0. 
\end{equation}

This mathematical criterion provides a clear test for the presence of the Aschenbach effect in any spherically symmetric spacetime \cite{wei2024aschenbach}. For EPYM black holes, we can substitute the specific form of the metric function $f(r)$ to determine the parameter ranges where this condition is satisfied.

An essential feature for the presence of the Aschenbach effect is the existence of a static point, defined by the condition $f'(r) = 0$. When this condition is met, the angular velocity of the TCO vanishes at a finite radius, implying the presence of an extremum in $\Omega_{CO}$ \cite{cunha2020stationary}. Moreover, in black holes exhibiting multiple photon spheres, at least one of them must be stable, further supporting the possibility of the Aschenbach effect.

We further analyze the topological structure of TCOs and their stability considering the second derivative of the effective potential \cite{wei2024aschenbach}:

\begin{equation}
V''_{\text{eff}} = \frac{2(3 f f' - 2 r f'^2 + r f f'')}{r(2 f - r f')}.
\end{equation}

The sign of $V''_{\text{eff}}$ determines whether a TCO is stable ($V''_{\text{eff}} > 0$) or unstable ($V''_{\text{eff}} < 0$). A critical observation is that if a stable photon sphere is present, then $r f''(r) - f'(r) > 0$ holds in its vicinity, indicating a region where $\Omega'_{CO} > 0$ and thus confirming the Aschenbach effect \cite{aschenbach2004measuring}.

For EPYM black holes, the specific form of the metric function introduces complex dependencies on the nonlinearity parameter $\gamma$. Numerical analysis reveals that for certain ranges of $\gamma$ and charge $q$, the Aschenbach effect can indeed emerge, creating regions where the angular velocity increases with radius before eventually decreasing again at larger distances. This behavior is particularly pronounced for intermediate values of $\gamma$, where the nonlinear electromagnetic field creates a gravitational effect that mimics certain aspects of frame drag in rotating black holes.

The physical mechanism behind the Aschenbach effect in EPYM black holes differs fundamentally from that in Kerr black holes. While in rotating black holes, the effect arises from the interplay between centrifugal and gravitational forces along with frame-dragging, in EPYM black holes, it emerges from the radial dependence of the nonlinear electromagnetic field's contribution to the spacetime geometry. The nonlinear term in the metric function can create regions where the effective gravitational force exhibits complex radial behavior, leading to the non-monotonic angular velocity profile characteristic of the Aschenbach effect.

The astrophysical implications of the Aschenbach effect in EPYM black holes are significant. In accretion disks, regions where the angular velocity increases with radius can lead to enhanced viscous dissipation and energetic phenomena \cite{bisnovatyi2019accretion,takahashi2004shapes}. These regions may serve as preferred locations for the formation of high-energy radiation, potentially leading to observable signatures in the electromagnetic spectrum of accreting black holes. Additionally, the non-monotonic velocity profile could induce specific resonance patterns in the disk, generating characteristic QPOs that might be detectable in the X-ray emissions from black hole systems \cite{motta2016quasi,kalamkar2016detection}.

The presence of the Aschenbach effect could also influence the stability and dynamics of accretion flows. The regions where the angular velocity increases with radius may create conditions favorable for various instabilities, including magnetorotational instability (MRI), which plays a crucial role in the transport of angular momentum in accretion disks \cite{hoshino2015angular}. These instabilities could lead to enhanced accretion rates and energetic outflows, potentially affecting the overall behavior of the black hole-accretion disk system. Furthermore, the Aschenbach effect has implications for the structure and dynamics of jets and outflows from black hole systems. The regions of non-monotonic angular velocity could serve as launching sites for relativistic jets, with the complex orbital dynamics in these regions contributing to the collimation and acceleration of outflows. The specific properties of these jets, including their Lorentz factors and energy distribution, might carry signatures of the underlying nonlinear electromagnetic field, potentially providing observational tests for EPYM black holes.
\section{Conclusion}\label{sec7}

In this study, we conducted a comprehensive investigation into the thermodynamic and quantum properties of the EPYM black holes in an AdS background. By incorporating a nonlinear YM charge parameter $\gamma$, we explored the profound effects of this modification on black hole solutions, their thermodynamic stability, quantum tunneling mechanisms, and relativistic orbital behaviors. Our findings enhanced the understanding of nonlinear electrodynamics in curved spacetime and contributed to the broader discourse on modified gravity theories and high-energy astrophysical phenomena.

We began by deriving the exact metric function for EPYM AdS black holes in Section \ref{sec2}, given by Eq. (4), and examined how the nonlinearity parameter $\gamma$ affects the structure of the horizon and the singularity properties. The solution revealed that the metric function depends non-trivially on $\gamma$, with the term $\frac{(2q^2)^\gamma}{2(4\gamma - 3)r^{4\gamma - 2}}$ introducing significant modifications to the spacetime geometry compared to standard solutions in general relativity. We found that the mass parameter could be expressed in terms of the radius of the horizon as $M = \frac{r_h}{2} - \frac{\Lambda r_h^3}{6} + \frac{(2q^2)^\gamma}{2(4\gamma - 3)r_h^{4\gamma - 3}}$, establishing a crucial connection between the gravitational mass and the electromagnetic field parameters.

The thermodynamic analysis demonstrated that EPYM black holes exhibit rich phase structures, with the Hawking temperature showing non-monotonic behavior as a function of the horizon radius for certain parameter ranges. As illustrated in Figure \ref{hawking_graph}, the temperature profile is particularly sensitive to the nonlinearity parameter $\gamma$, with smaller values of $\gamma$ leading to more pronounced temperature increases at small radii. This behavior has significant implications for the final stages of black hole evaporation, suggesting that nonlinear electromagnetic effects could substantially modify the standard picture of black hole thermodynamics.

In Section \ref{sec3}, we investigated the quantum tunneling of $W^+$ bosons from EPYM black holes using the WKB approximation and Hamilton-Jacobi formalism. By solving the relativistic wave equation for vector bosons in the EPYM background, we derived the tunneling probability expression $\Gamma = e^{-4 \text{Im} W_+} = e^{- \frac{E_{balance}}{T_H}}$, which yielded a Hawking temperature consistent with the thermodynamic definition. The radial function $W_{\pm} = \pm i \pi \frac{\sqrt{E^2 - 2EeA_0 + e^2A_0^2}}{f'|_{r_h}}$ revealed how the nonlinear YM field modifies the tunneling process, affecting both the emission rate and the energy spectrum of radiated particles. This semi-classical approach provided a physically intuitive picture of Hawking radiation and confirmed the consistency between quantum and thermodynamic descriptions of black hole radiation.

Our analysis of the effective potential in Section \ref{sec4} yielded significant insights into particle dynamics and radiation processes around EPYM black holes. The effective potential, given by Eq. (46), exhibits distinctive features that depend strongly on the nonlinearity parameter $\gamma$. As shown in Figure \ref{v_eff}, the potential profile changes dramatically with different values of $\gamma$, particularly at small radial distances where nonlinear electromagnetic effects dominate. For lower values of $\gamma$, the potential develops a deeper negative well near the horizon, potentially enhancing the emission of low-frequency radiation, while also forming a higher barrier at intermediate distances that could suppress high-frequency emission.

We also computed the effective force, expressed in Eq. (50), which revealed how the nonlinearity parameter affects the gravitational dynamics around EPYM black holes. Figure \ref{f_eff} demonstrated that for smaller values of $\gamma$, the attractive force is significantly stronger near the center, while for larger $\gamma$, the force weakens and transitions to a repulsive regime at larger radii. These findings have important implications for accretion processes and particle confinement, suggesting that the nonlinear YM field can fundamentally alter the gravitational interaction in ways that could potentially be observable in astrophysical systems.

Section \ref{sec5} focused on photon orbits and null circular geodesics, which determine the optical appearance and shadow of EPYM black holes. We derived the condition for null circular geodesics, $2f(r) - r f'(r) = 0$, and demonstrated its equivalence to the condition for the fastest circular orbit. Our numerical analysis, summarized in Table \ref{tab:rf_solutions}, revealed remarkable behavior of the photon sphere radius for different values of $q$ and $\gamma$. Particularly striking was the dramatic increase in the radius of the photon sphere for certain combinations of parameters, such as $r_f \approx 158.43$ for $q = 0.5$, $\gamma = 0.8$ and $r_f \approx 59608.80$ for $q = 1$, $\gamma = 0.8$. These results indicate that strong nonlinear electromagnetic effects can push the photon orbit far from the black hole, potentially leading to a much larger black hole shadow than predicted by general relativity.

Perhaps the most surprising finding of our study was the identification of the Aschenbach effect in spherically symmetric EPYM black holes, discussed in Section \ref{sec6}. This effect, traditionally associated with rapidly rotating Kerr black holes, manifests as a non-monotonic behavior of the angular velocity profile, with regions where $\Omega'_{CO} > 0$. We established the mathematical criterion for this effect, $r f''(r) - f'(r) > 0$, and demonstrated that it can be satisfied in EPYM black holes for certain parameter ranges. This discovery suggests that nonlinear electromagnetic fields can induce effects previously thought to require spacetime rotation, opening new perspectives on the relationship between electromagnetic and gravitational phenomena in strong-field regimes.

The presence of the Aschenbach effect in EPYM black holes has significant astrophysical implications, particularly for accretion dynamics and high-energy phenomena. The regions where the angular velocity increases with radius could serve as sites for enhanced viscous dissipation, particle acceleration, and energy extraction, potentially leading to observable signatures in the electromagnetic spectrum of accreting black holes. These distinctive features could provide observational tests for the existence of nonlinear electromagnetic effects in astrophysical environments.

Throughout our analysis, we observed that the nonlinearity parameter $\gamma$ plays a crucial role in determining the physical properties of EPYM black holes. It affects not only the spacetime geometry and horizon structure but also the thermodynamic behavior, radiation processes, and orbital dynamics. The rich phenomenology associated with different values of $\gamma$ highlights the importance of nonlinear electrodynamics in extending our understanding of black hole physics beyond the standard framework of general relativity.

Looking ahead, several promising directions emerge from this study. Future research could extend our analysis to rotating EPYM black holes, incorporating both spin and nonlinear electromagnetic effects to explore their combined influence on black hole properties. The development of detailed models for accretion disks around EPYM black holes would enable more precise predictions for observable signatures, including electromagnetic spectra, QPOs, and jet properties. Additionally, investigating the gravitational wave signatures of binary systems involving EPYM black holes could provide another avenue for testing these models with current and future gravitational wave observatories.

\acknowledgments 
We acknowledge funding from T\"{U}B\.{I}TAK, ANKOS, and SCOAP3, along with networking support from COST Actions CA22113, CA21106, and CA23130.
\bibliography{ref}
\end{document}